\documentclass[]{spie}  

\usepackage{amsmath,amsfonts,amssymb}
\usepackage{graphicx}
\usepackage[colorlinks=true, allcolors=blue]{hyperref}
\usepackage{float}
\title{New starting point registration method for tagged MRI tongue motion estimation}

\author[a]{Jinglun Yu}
\author[a]{Muhan Shao}
\author[a]{Zhangxing Bian}
\author[b]{Xiao Liang}
\author[b]{\\Jiachen Zhuo}
\author[c]{Maureen Stone}
\author[a]{Jerry L.~Prince}
\affil[a]{Department of Electrical and Computer Engineering, Johns~Hopkins~University,~Baltimore,~MD~21218,~USA}
\affil[b]{Department of Diagnostic Radiology and Nuclear Medicine, University~of~Maryland~School~of~Medicine,~Baltimore,~MD~21201,~USA}
\affil[c]{Department of Neural and Pain Sciences, University~of~Maryland~Dental~School,~Baltimore,~MD~21201,~USA}
\authorinfo{Further author information: (Send correspondence to Jinglun Yu)\\Jinglun Yu: E-mail: jyu146@jhu.edu\\}

\begin{document} 
\maketitle
\begin{abstract}
Analysis of tongue motion has been proven useful in gaining a better understanding of speech and swallowing disorders. Tagged magnetic resonance imaging~(MRI) has been used to image tongue motion, and the harmonic phase processing~(HARP) method has been used to compute 3D motion from these images. However, HARP can fail with large motions due to so-called tag~(or phase) jumping, yielding highly inaccurate results. The phase vector incompressible registration algorithm~(PVIRA) was developed using the HARP framework to yield smooth, incompressible, and diffeomorphic motion fields, but it can also suffer from tag jumping.  In this paper, we propose a new method to avoid tag jumping occurring in the later frames of tagged MR image sequences. The new approach uses PVIRA between successive time frames and then adds their stationary velocity fields to yield a starting point from which to initialize a final PVIRA stage between troublesome frames.  We demonstrate on multiple data sets that this method avoids tag jumping and produces superior motion estimates compared with existing methods.  
\end{abstract}


\section{INTRODUCTION}
\label{sec:intro} 
The tongue is involved in several vital activities including speaking and swallowing~\cite{3-TongueStructure}. Since it has no bones, the tongue changes its shape using only its own muscles. An accurate measure of tongue deformations during such activities is essential in gaining a better understanding of these vital activities and the tongue's role in them. Since the volume change of the tongue during these activities is negligible, its internal motion field can be assumed to be diffeomorphic and incompressible~\cite{1-PVIRA}. 

\begin{figure} [H]
   \begin{center}
   \begin{tabular}{c} 
   \includegraphics[height=3.2cm]{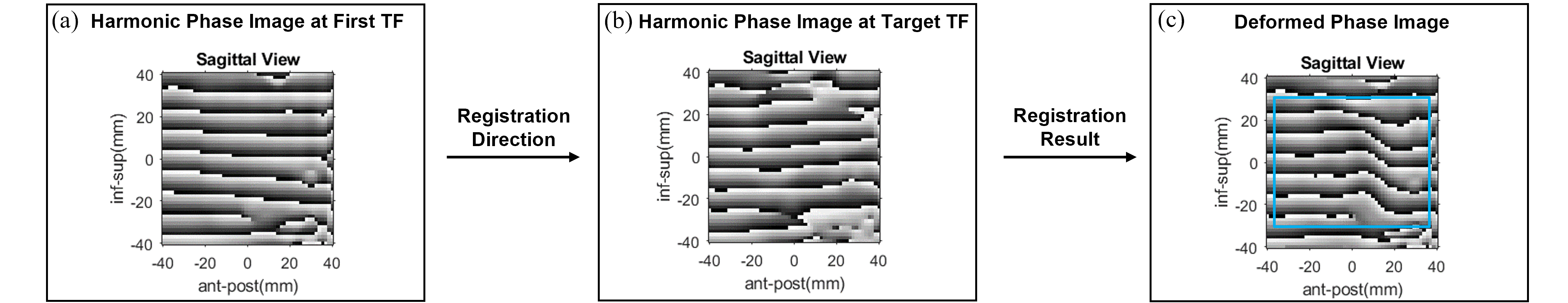}
   \end{tabular}
   \end{center}
   \caption[fig:1] 
   { \label{fig:1} 
   Example of the tag jumping phenomenon for tagged MRI. \textbf{(a)}~Harmonic phase image at first time frame. \textbf{(b)}~Harmonic phase image at target time frame. \textbf{(c)}~Deformed phase image. Blue box: tag jumping area.}
   \end{figure}

The phase vector incompressible registration algorithm~(PVIRA)\cite{1-PVIRA} can be used to reconstruct the tongue motion from the tagged MRI data\cite{4-TaggedMRI}. This method uses harmonic phase volumes obtained by HARP filtering~\cite{5-HARPVijay} as input to the iLogdemons~\cite{6-iLogDemons} algorithm to yield incompressible motion and an inverse motion field. This is done by computing the stationary velocity field iteratively starting from an initial zero velocity field. However, when the motion between the two input images is large, the closest matching phase may come from a different sinusoidal cycle---i.e., tag jumping---leading to grossly erroneous motion estimates. Figure~\ref{fig:1} shows a set of tagged MRI phase images that are affected by tag jumping. To avoid tag jumping, existing methods compute incremental motion fields between frames that are closer together in time and then compose these fields, which leads to accumulating errors. 

In this paper, we propose a method that uses a combined stationary velocity field to replace the zero initialization in the velocity computation iteration of the iLogDemons algorithm. This approach gives the velocity iteration a new starting point that is close to the desired velocity field,  which mitigates abrupt motion changes between two registration images. The experimental results demonstrate that this approach can avoid gross motion errors due to tag jumping and therefore produce more accurate motion estimates. 

\section{METHODS}
\subsection{Data Collection}
\label{sec:title}
The data sets in this paper consist of four healthy subjects who are native speakers of American English. The speech task was to pronounce /athing/. The data sets are 3D tagged images with 6 mm slice thickness and $1.875 \times 1.875$~mm in-plane resolution captured using 26 time-frames~(TFs) in one second. The tagged image volumes are converted to three HARP phase image volumes for motion estimation using PVIRA~\cite{1-PVIRA} as described below.  

\subsection{Direct and Incremental Registration Method}
\begin{figure}
   \begin{center}
   \begin{tabular}{c} 
   \includegraphics[height=5.2cm]{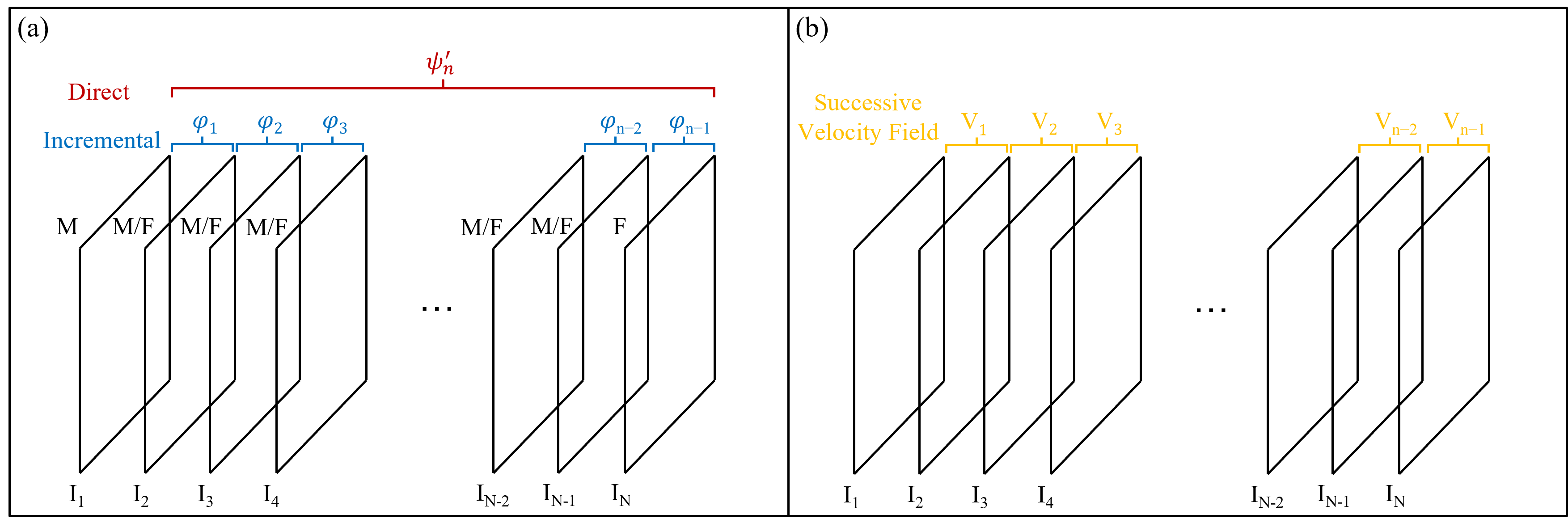}
   \end{tabular}
   \end{center}
   \caption[fig:2] 
   { \label{fig:2} 
   \textbf{(a)}~The schematic diagram of direct and incremental~(red and blue) registration methods. M: moving image, F: fixed image. I: each TF phase image. \textbf{(b)}~The schematic diagram of the successive velocity field between each TF.}
   \vspace{0.6cm}
   \end{figure}

The direct and incremental approaches are two ways to use PVIRA~\cite{1-PVIRA} to estimate 3D motion. The direct method always uses the first phase image $I_1$ as the moving image and the $n$-th phase image $I_n$ as the fixed image, as shown in Fig.~\ref{fig:2}(a). The resulting motion estimates from the first TF to all other TFs are:
\begin{equation}
\psi'_n, \quad n = 2,\ldots,N
\end{equation}
The incremental method uses each successive phase image as the moving image and the next image in the sequence as the fixed image, as shown in Fig.~\ref{fig:2}(a). This yields a sequence of deformations, $\varphi_i$, $i=1,\dots, N-1$.  The resulting motion estimates from the first TF to all other TFs are: 
\begin{equation}
\psi''_n = \varphi_{n-1} \circ \varphi_{n-2} \cdots \varphi_{2} \circ \varphi_1 , \quad n = 2,\ldots, N
\end{equation}
where $\circ$ is composition. If the motion is too large between the $1^{\text{st}}$ and $n^{\text{th}}$ TF, the direct method may suffer from tag jumping.  The incremental method avoids this by composing smaller deformations, but suffers from accumulating errors over time.   

   \begin{figure}[!tb]
   \begin{center}
   \begin{tabular}{c} 
   \includegraphics[height=8.3cm]{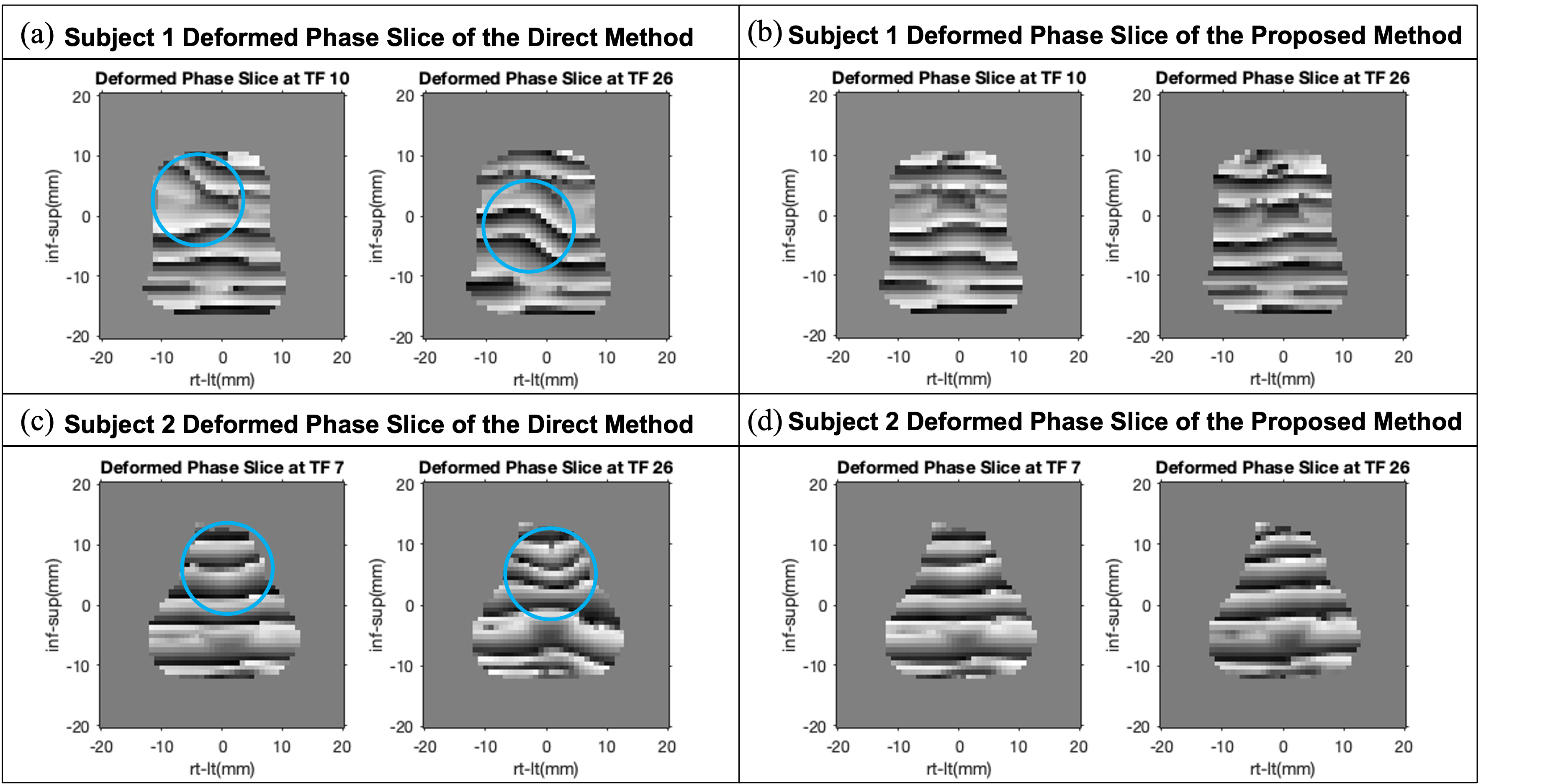}
   \hspace{-10mm}
   \end{tabular}
   \end{center}
   \caption[fig:3] 
   { \label{fig:3} 
   \textbf{(a)}~Subject~1 deformed phase slice of the direct method. \textbf{(b)}~Subject~1 deformed phase slice of the proposed method.
   \textbf{(c)}~Subject~2 deformed phase slice of the direct method.
   \textbf{(d)}~Subject~2 deformed phase slice of the proposed method.
   Blue circles: tag jumping area. }
   \end{figure} 
   
\subsection{New Starting Point Method}
The tag jumping problem is caused by a deformation that is too large.  In this case, PVIRA latches onto matching HARP phases that correspond to the wrong sinusoidal tag period.  To address this, PVIRA simply needs a closer starting point.  PVIRA is based on the iLogDemons algorithm~\cite{6-iLogDemons}, which estimates a stationary velocity field in the log domain followed by fast exponential map to produce a displacement. To produce a closer starting point, as shown in Fig.~\ref{fig:2}(b), we compose the incremental deformations by adding their stationary velocity fields as follows:
\begin{equation}
V'_n= V_{n-1} + V_{n-2} + \cdots + V_2 + V_1, 
\qquad n = 2, \ldots, N
\end{equation}
which forms a new starting stationary velocity field for a modified PVIRA method that takes $I_1$, $I_{n}$, and $V'_n$ as input. PVIRA uses $V'_n$ rather than zero as the starting stationary velocity field, but is otherwise unchanged.  
This approach yields a new deformation (i.e., resulting motion estimate) from $I_1$ to $I_n$, which we denote by $\psi'''_n$.

\section{EXPERIMENTS AND RESULTS}
\label{sec:sections}
\begin{figure}[H]
   \begin{center}
   \begin{tabular}{c} 
   \hspace{-0.2cm}
   \includegraphics[height=7.6 cm]{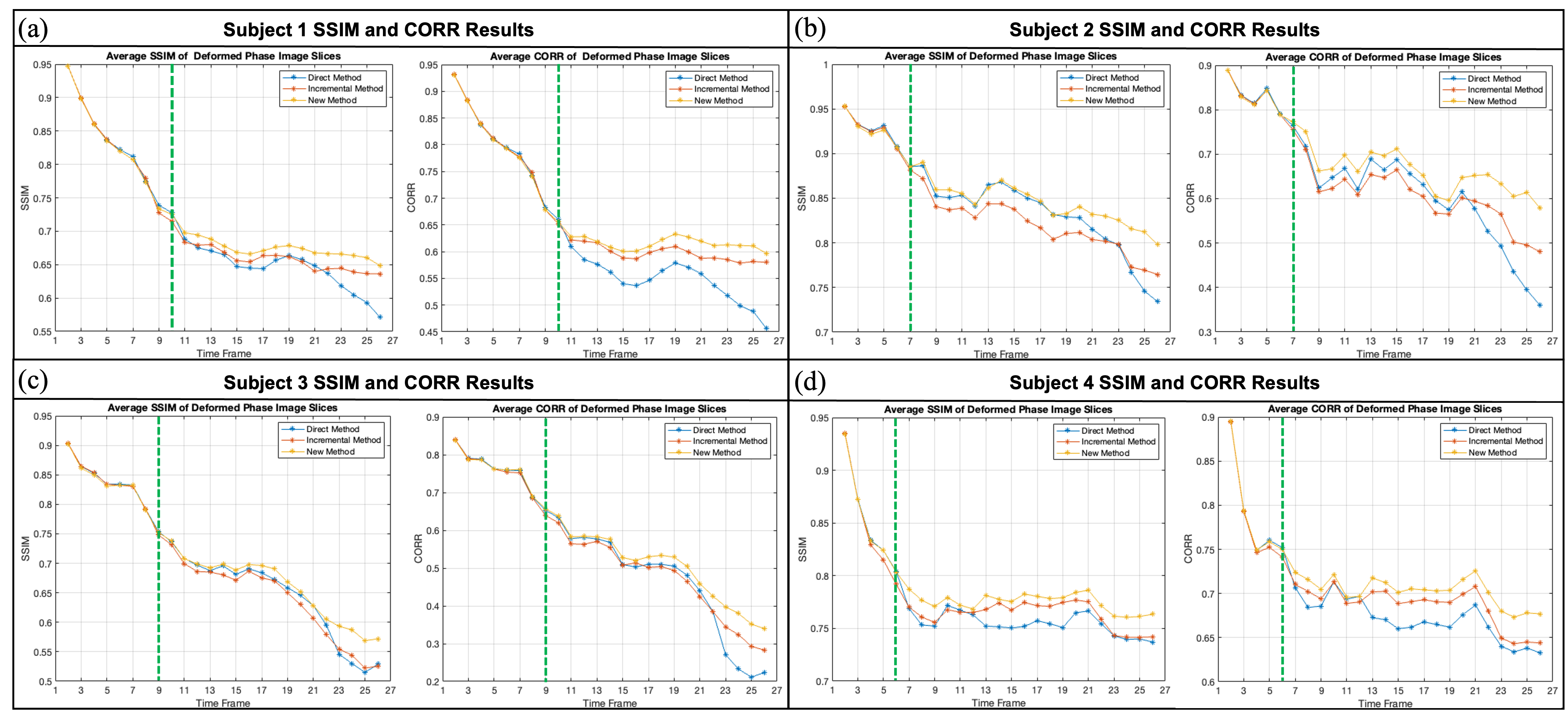}
   \end{tabular}
   \end{center}
   \caption[fig:4] 
   { \label{fig:4} 
   Average SSIM and CORR of deformed phase image slices. 
   \textbf{(a)}~SSIM and CORR results of Subject~1. 
   \textbf{(b)}~SSIM and CORR results of Subject~2.
   \textbf{(c)}~SSIM and CORR results of Subject~3. 
   \textbf{(d)}~SSIM and CORR results of Subject~4.
   Green dashed lines: starting time frame of tag jumping.} 
   \end{figure}

\begin{figure}[H]
   \begin{center}
   \begin{tabular}{c} 
   \hspace{-0.2cm}
   \includegraphics[height=7.6 cm]{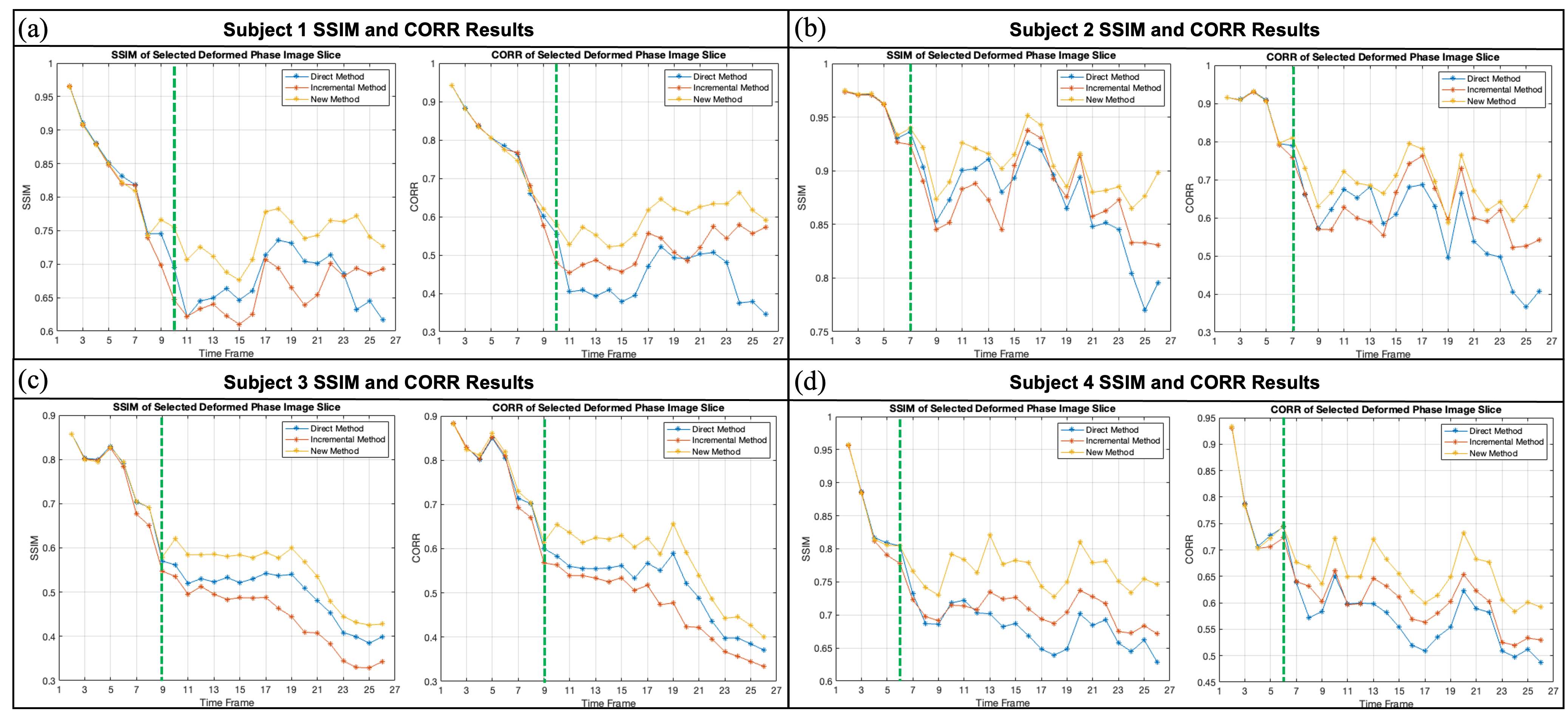}
   \end{tabular}
   \end{center}
   \caption[fig:5] 
   { \label{fig:5} 
   SSIM and CORR of selected deformed phase image slice. 
   \textbf{(a)}~SSIM and CORR results of Subject~1. 
   \textbf{(b)}~SSIM and CORR results of Subject~2.
   \textbf{(c)}~SSIM and CORR results of Subject~3. 
   \textbf{(d)}~SSIM and CORR results of Subject~4.
   Green dashed lines: starting time frame of tag jumping.}
   \end{figure}

The performance of the proposed method was compared to two state-of-the-art registration methods: the direct method\cite{1-PVIRA} and the incremental method. The performance evaluation is based on both the ability to avoid tag jumping and the motion estimate accuracy. For quantitative accuracy evaluation, we use the harmonic phase images generated by the HARP filter\cite{7-Osman} as the ground truth of the deformed phases. The experiments warp the motion field results obtained from each method to the first TF harmonic phase image to form the deformed phase results at $n$-th TF using the Advanced Normalization Tools~(ANTs)~\cite{8-ANTs}: 
\begin{equation}
\hat{I}_n = \psi_n \circ I_1,
\qquad n = 2, \ldots, N
\end{equation}
Manual review of each method’s deformed phase image slices can demonstrate the ability to avoid tag jumping. The slice example from Subject 1 and Subject 2 in Fig.~\ref{fig:3} shows that the direct method is suffering from tag jumping between TF~10 and TF~26 for Subject 1, TF~7 and TF~26 for Subject 2, while the proposed method avoids this problem in corresponding TFs. The structural similarity index measure (SSIM)~\cite{2-SSIM} and the correlation coefficient (CORR)~\cite{9-CORR} between the harmonic and deformed phase slices of each method are computed to evaluate motion estimation accuracy. Figure~\ref{fig:4} compares the average SSIM and CORR over all deformed phase image slices at each TF among three methods. When there is no tag jumping, the proposed method has the same performance with the other two methods. During TFs~10--26 of Subject~1, TFs~7--26 of Subject~2, TFs~9--26 of Subject~3, and TFs~6--26 of Subject~4, where tag jumping occurs, the new starting point method yields consistently higher similarity with ground truth images than the direct and incremental registration methods, indicating correction of phase jumping while maintaining more accurate motion estimates. To highlight the substantial improvement of motion estimates in the deformed phase images with tag jumping, we selected the most severely affected slice at the tag jumping TFs as shown in Fig.~\ref{fig:5}. In the tag jumping TFs with respect to all subjects, the mean SSIM increases by 5.55$\%$ and 7.32$\%$  and the mean CORR increases by 14.84$\%$ and 9.12$\%$ for the direct and incremental methods, respectively. 

\section{DISCUSSION AND CONCLUSION}
In this paper, we proposed a new registration method that can construct the motion estimate from tagged MRI data while eliminating the tag jumping problem. A new velocity initialization was used in a modified PVIRA algorithm to obtain a starting point that is closer to the correct final result. The experiments show that the proposed method outperforms the two comparison motion estimation methods, the direct and incremental PVIRA methods. The new approach reduces the possibility of tag jumping and incremental error accumulation assuming that intermediate time frames with smaller incremental motions are available. Future work will include replacing the generation of the deformed phase images by ANTs with PVIRA.
   
\acknowledgments 
 This work was supported by National Institute on Deafness and Other Communication Disorders Grant R01-DC014717 (PI: Jerry L. Prince). The study was conducted at University of Maryland School of Medicine Center for Innovative Biomedical Resources, Translational Research in Imaging(CTRIM).
     
\bibliography{report} 
\bibliographystyle{spiebib} 
   
\end{document}